\documentclass[onecolumn,prb,showpacs,showkeys,amsmath,amssymb,floatfix]{revtex4}

\usepackage{graphicx}
\usepackage{dcolumn}
\usepackage{bm}

\begin{document}

\title{Electron spin relaxation in GaAs quantum dot systems -- The role of the hyperfine interaction}

\author{Johannes Voss}
\email{johannes.voss@fysik.dtu.dk}
\altaffiliation[Present address: ]{%
Ris\o{} National Laboratory, %
Materials Research Department, %
Technical University of Denmark, %
Frederiksborgborgvej 399, DK-4000 Roskilde
}
\affiliation{%
Universit\"{a}t Hamburg\\
1.\ Institut f\"{u}r Theoretische Physik\\
Jungiusstra\ss{}e 9, D-20355 Hamburg
}

\author{Daniela Pfannkuche}
\email{daniela.pfannkuche@physik.uni-hamburg.de}
\affiliation{%
Universit\"{a}t Hamburg\\
1.\ Institut f\"{u}r Theoretische Physik\\
Jungiusstra\ss{}e 9, D-20355 Hamburg
}

\date{December 14, 2007}

\begin{abstract}
We present numerical results for electron spin relaxation rates for single and laterally coupled double GaAs quantum dots in a perpendicular magnetic field. As source of spin relaxation we consider hyperfine interaction with the nuclear spins in the GaAs substrate. Due to the differences in the energy scales of the nuclear and electronic Zeeman energies, the phonon bath system has to be taken into account for energy dissipation. The corresponding transition rates of second order show strong dependencies on correlations between the electrons and the electronic energy differences, and hence on the magnetic field. For a highly asymmetric double dot we have found a relatively low second order electron spin relaxation rate for a wide range of magnetic fields.
\end{abstract}

\pacs{73.21.La, 76.60.Es}
\keywords{electron spin relaxation; laterally coupled double quantum dot; hyperfine interaction; quantum gate; asymmetric double dot; spin dynamics}

\maketitle

\section{Introduction}

Strong efforts have been placed on testing and developing semiconductor quantum dots for quantum computing applications.\cite{burkard,hansonprl94,hanson:1217} For the implementation of a device that allows for operation on two spatially separated electron spins, {\it i.e.}~a quantum gate, Burkard, Loss, and DiVincenzo have proposed two laterally coupled quantum dots containing one electron each.\cite{burkard} Such a quantum dot system can be designed with metallic gates on a hetero structure. With variation of the gate voltages a controlled, time-dependent inter-dot coupling can be achieved. This is an advantage compared to vertically coupled quantum dots.

For (quantum) computation, the states representing the information should at least be stable during the calculation and the read-out afterwards. Hence it is very important and interesting to study the mechanisms of spin relaxation for these quantum dot systems.\cite{borhani:155311,hanson:050502,climente:081303,hu:100501,deng,khaetskiiprb67,merkulov,schliemann,abalmassov,khaetskiiprb64,bulaev,holleitner:075204,braun:116601,meunier:126601} For magnetic fields of the order of $1 \textrm{T}$ and above, spin-orbit coupling is considered the most effective mechanism of electron spin relaxation in GaAs quantum dots,\cite{bulaev,khaetskiiprb67} while for smaller magnetic fields the hyperfine interaction with the nuclear spins in the GaAs substrate is assumed to be dominant.\cite{erlingsson} Johnson \emph{et al.}\cite{johnson} have shown that up to decay times of $1 {\rm ms}$, the triplet-singlet relaxation of two electron spins in a GaAs double quantum dot in a small perpendicular magnetic field is governed by hyperfine interaction with the nuclear spins. For a numerical study of relaxation of correlated electron spins due to spin orbit coupling see [\onlinecite{florescu:045304}].

Erlingsson\cite{erlingsson} has calculated electron spin lifetimes for two non-interacting electrons in a single quantum dot due to a second order hyperfine/electron-phonon interaction process. We extend his approach to interacting electrons and to more complicated quantum dot structures, {\it i.e.}~(asymmetric) laterally coupled double dots, allowing for a tayloring of electron spin relaxation rates.

\section{Theory}

\subsection{Transition rate formalism}

The electron spin relaxation rate due to interaction between the electrons in the quantum dot and the nuclear spin and phonon bath systems is calculated via Fermi's golden rule. At nonzero magnetic fields, the nuclear Zeeman energy is much smaller than the electronic Zeeman energy. Furthermore, for $S^2$ transitions also orbital energy differences need to be absorbed. Except for energetic crossings between the initial and final electronic states of the spin flip, the law of energy conservation forbids these ``hyperfine-only'' relaxations. The electron-phonon interaction in turn cannot directly attack the electronic spin degree of freedom. Yet, relaxation is possible in second order via a combined hyperfine and electron-phonon interaction process. Thermally averaging over the initial bath states, the transition rate from the electronic state $| i \rangle$ to the state $| f \rangle$ is\cite{blum,sakurai}
\begin{eqnarray}
\Gamma_{i \rightarrow f} & = & \frac{2 \pi}{\hbar} \, \delta(\epsilon_i - \epsilon_f - \hbar \omega) \sum_{\mu \mu' N N'} P(\mu) P(N) \nonumber\\
 & & \times \left| \sum_{\ell \neq i,f} \frac{\langle f ; \mu' | \hat V_{\rm HF} | \ell ; \mu \rangle \, \langle \ell ; N' | \hat V_{\rm PH} | i  ; N \rangle }{\epsilon_f - \epsilon_{\ell}} \right. \nonumber\\
 & & {\left. + \frac{\langle f ; N' | \hat V_{\rm PH} | \ell ; N \rangle \, \langle \ell ; \mu' | \hat V_{\rm HF} | i  ; \mu \rangle }{\epsilon_i - \epsilon_{\ell}} \right|}^2 .\label{goldenrule}
\end{eqnarray}
$\epsilon_{\ell}$ is the energy of the electronic state $| \ell \rangle$. $\hbar \omega$ is the energy of the phonon emitted during the relaxation process. The nuclear spin flip energy is neglected in the energy differences in Eq.~(\ref{goldenrule}), because it is much smaller than the corresponding electronic energies. $P(\mu)$ and $P(N)$ are the probabilities to initially find the nuclear system and the phonon bath system in the states $| \mu \rangle$ and $| N \rangle$, respectively. Eq.~(\ref{goldenrule}) implies that there initially are no correlations between the electronic subsystem and the bath systems.

The initial state $| i \rangle$ and the final state $| f \rangle$ are the lowest energy states in the singlet/triplet sector. 
Eq.~(\ref{goldenrule}) diverges for the case of an energetic crossing between the initial state $| i \rangle$ and an intermediate state $| \ell \rangle$ with the same spin as the final state $| f \rangle$ (or the corresponding case for a crossing of final and intermediate state). The resulting vanishing energy denominators would need a thorough treatment.\cite{schiff} For the parameter ranges we studied, these types of crossings did not appear. Crossings between intermediate states do not cause pronounced effects on the second order transition rate, because the contributions to the rate of the corresponding transition elements are continuously weighted with the energy differences to the initial and final state, respectively.

The Markovian approximation involved in the usage of Fermi's golden rule generally is valid for the interaction with the phonon bath due to short phonon lifetimes\cite{romanov} compared to typical electronic timescales in semiconductor quantum dots. It is known that the system of nuclear spins can be polarized by the hyperfine interaction with a 2DEG electron gas.\cite{dobers,ono} In our case, however, the number of electrons is small. We can therefore assume that the nuclear system is not polarized for all times and the few dot electrons will not carry out enough collisions to polarize the nuclear system.

\subsection{The hyperfine interaction}

For an $s$-band electron the hyperfine interaction is of Fermi contact type:\cite{slichter}
\begin{equation}
\hat H^{(1)}_{\rm HF} = A_k \, \hat {\bm S}^{(1)} \cdot \hat {\bm I}_k \, \delta({\bm r} - {\bm R}_k) , \label{vcont}
\end{equation}
where $\hat {\bm S}^{(1)}$ and $\hat {\bm I}_k$ are the single-electron (superscript $(1)$) spin operator and nuclear spin operator of the $k$th nucleus, respectively. ${\bm r}$ and ${\bm R}_k$ are the corresponding positions. For GaAs consisting of isotopes with $I=3/2$ but different nuclear gyromagnetic ratios, we consider an average hyperfine constant $A$. From electron spin resonance data for conduction electrons in a GaAs layer of a GaAs/Al${}_x$Ga${}_{1-x}$As heterostructure,\cite{vitkalov} an approximate value for $A$ of $-2.5 \cdot 10^{-3} {\rm m}^3 {\rm eV}^{-1} {\rm s}^{-2}$ can be calculated.

Projecting (\ref{vcont}) on the electronic orbitals $| n \rangle$, we notice that the hyperfine interaction is modulated by the orbitals wavefunctions at the positions of the nuclei:
\begin{eqnarray}
\hat H^{(1)}_{\rm HF} & = & A \, \hat {\bm S}^{(1)} \cdot \hat {\bm I}_k \left( \sum_n | \langle {\bm R}_k | n \rangle |^2 \, | n \rangle \langle n | \right. \nonumber\\
 & & \left. + \sum_{n n'} \langle n' | {\bm R}_k \rangle \langle {\bm R}_k | n \rangle \, | n' \rangle \langle n | \right) . \label{diagandndiag}
\end{eqnarray}
The first sum, diagonal in the orbitals, only shifts the electronic eigenergies. We will take into account these diagonal terms by an effective magnetic field, which in addition to the external magnetic field shifts the electronic energies via a Zeeman term.\cite{paget,erlingsson,merkulov} The effective magnetic field for an electron with orbital quantum number $n$ is calculated with the thermal meanvalue
\begin{equation}
{\bm B}_{\rm eff} = {\left\langle \frac{\hbar A}{g_e^{*} \mu_B} \sum_k \hat {\bm I}_k | \langle {\bm R}_k | n \rangle |^2 \right\rangle}_T ,
\end{equation}
with the effective electronic Land\'e factor $g_e^{*}$. Each ${\bm I}_k$ has the same thermal expectation value for non-interacting nuclear spins:
\begin{eqnarray}
{\langle \hat I_z \rangle}_T & = & \sum_{i=0}^3 \hbar \left(i-\frac{3}{2}\right) \exp\!\left( \frac{-\overline{\gamma_N} B_z \hbar \left(i-\frac{3}{2}\right)}{k_{\rm B} T} \right) \nonumber\\
& & \times {\left[ \sum_{k=0}^3 \exp\!\left( \frac{-\overline{\gamma_N} B_z \hbar \left(k-\frac{3}{2}\right)}{k_B T} \right) \right]}^{-1} \label{expiz} ,
\end{eqnarray}
where the magnetic field is pointing in $z$-direction. $\overline\gamma_N \approx 58.5 \cdot 10^6 {{\rm s}^{-1}\,{\rm T}^{-1}}$ is an average nuclear gyromagnetic ratio. $k_{\rm B}$ is the Boltzmann constant.
Since the orbitals can be assumed to vary little on the nuclear distances, we can approximate the sum over the nuclei by a three-dimensional integral and have
\begin{equation}
{\bm B}_{\rm eff} \approx {\langle {\bm I} \rangle}_T \frac{8 \hbar A}{g_e^{*} \mu_B a_0^3},
\end{equation}
with the lattice constant $a_0$. We have dropped the index $k$ since all nuclear spin expectation values are equal. Important for the validity of this semiclassical effective field are a large number of nuclei and an unpolarized nuclear spin bath.\cite{erlingsson} We notice, that ${\bm B}_{\rm eff}$ does not depend on the orbital quantum number of the electron.

We treat the nondiagonal terms in Eq.~(\ref{diagandndiag}) perturbatively, the electronic Zeeman energy scale being much smaller than the orbital energy differences. Introducing the creation and annihilation operators $\hat a^+_{n \sigma}$ and $\hat a^{~}_{n \sigma}$ for an electron in orbital $n$ with spin quantum number $\sigma \in \{ \uparrow , \downarrow \}$, we have
\begin{eqnarray}
\hat V_{\rm HF} & = & A \sum_k {\sum_{n n' \sigma \sigma'}}^{\!\!\!\prime} \left( \sum_{\alpha \in \{x , y , z\}} \hat I_{k,\alpha} \langle \sigma | \hat S^{(1)}_{\alpha} | \sigma' \rangle \right) \nonumber\\
& & \times \langle n | {\bm R}_k \rangle \langle {\bm R}_k | n' \rangle \, \hat a^+_{n \sigma} \hat a^{~}_{n' \sigma'} \quad . \label{vhf}
\end{eqnarray}
With the prime on the sum over the electronic quantum numbers we denote that the sum runs over all quantum numbers except for which $n = n'$ and $\sigma = \sigma'$. These diagonal terms are included in the effective magnetic field. The sum over the nuclei is again approximated by a three-dimensional integral.

Evaluating the absolute square in Eq.~(\ref{goldenrule}), the hyperfine interaction operator (\ref{vhf}) appears in products of matrix elements of the form
\begin{equation}
\sum_{\mu} P(\mu) \langle a ; \mu | \hat V_{\rm HF} | b \rangle \langle c | \hat V_{\rm HF} | d ; \mu \rangle\label{vhffourorb}
\end{equation}
($| a \rangle$, $| b \rangle$, $| c \rangle$, and $| d \rangle$ are multi-electron states). Following Erlingsson,\cite{erlingsson} we redefine the nuclear spin bath state vector to only consist of the fluctuating pair $\hat {\bm I}_k , \hat {\bm I }_{k'}$:
\begin{equation}
| \mu \rangle \rightarrow | \mu_k \rangle \otimes | \mu_{k'} \rangle ,
\end{equation}
assuming that the nuclear system always is unpolarized, since few electrons are not sufficient for a polarization.
With
\begin{equation}
\delta \hat I_{\alpha \in \{ x , y , z \}} = \hat I_{\alpha} - {\langle \hat I_{\alpha} \rangle}_T
\end{equation}
we write the fluctuations in the hyperfine interaction with the $k$th nucleus due to nuclear spin fluctuations as
\begin{eqnarray}
\delta \hat V^{(k)}_{\rm HF} & = & A {\sum_{n n' \sigma \sigma'}}^{\!\!\!\prime} \, \sum_{\alpha \in \{ x , y , z \}} \delta \hat I_{k,\alpha} \langle \sigma | \hat S^{(1)}_{\alpha} | \sigma' \rangle \nonumber\\
 & & \langle n | {\bm R}_k \rangle \langle {\bm R}_k | n' \rangle \, \hat a^+_{n \sigma} \hat a^{~}_{n' \sigma'} .
\end{eqnarray}
The sum over the initial nuclear spin bath states is replaced by a sum over all pairs of fluctuating nuclear spins:
\begin{eqnarray}
\hspace{-4.5ex}&&\sum_{\mu} P(\mu) \langle a ; \mu | \hat V_{\rm HF} | b \rangle \langle c | \hat V_{\rm HF} | d ; \mu \rangle \nonumber\\
\hspace{-4.5ex}&\rightarrow&\sum_k \sum_{\mu_k} P(\mu_k) \, \langle \mu_k | \, \langle a | \hat V_{\rm HF} | b \rangle \langle c | \hat V_{\rm HF} | d \rangle \, | \mu_k \rangle . \label{pairsum}
\end{eqnarray}
$P(\mu_k)$ denotes the Boltzmann weight of a single (nuclear) spin in a static magnetic field (cf.~Eq.~(\ref{expiz})). Fluctuations among components of different nuclear spins are neglected, since the dipolar interaction energies are much smaller than the nuclear Zeeman energies.

The correlations in Eq.~(\ref{pairsum}) for different components of the same spin are given by $(\alpha,\beta,\gamma \in \{ x , y , z \})$
\begin{equation}
{\langle \delta \hat I_{\alpha} \delta \hat I_{\beta \neq \alpha} \rangle}_T = \sum_{\gamma} \epsilon_{\alpha \beta \gamma} i \hbar {\langle \hat I_{\gamma} \rangle}_T ,
\end{equation}
for the same components by
\begin{equation}
{\langle \delta \hat I_{\alpha} \delta \hat I_{\alpha} \rangle}_T = {\langle \hat I^2_{\alpha} \rangle}_T - {\langle \hat I_{\alpha} \rangle}^2_T .
\end{equation}

\subsection{The electron-phonon interaction}

The electron-phonon interaction enters the second order transition rate in form of the products
\begin{equation}
\sum_{N} P(N) \langle a ; N | \hat V_{\rm PH} | b \rangle \langle c | \hat V_{\rm PH} | d ; N \rangle \, \delta(\epsilon_i - \epsilon_f - \hbar \omega) \label{phondoppel}
\end{equation}
($| a \rangle$, $| b \rangle$, $| c \rangle$, and $| d \rangle$ are multi-electron states.) $N$ enumerates the phonons in mode $\omega$.
As we are studying electronic relaxation processes at low temperatures, we restrict the interaction to phonon emission. Hereby, the sum over the phonon bath states $| N \rangle$ times the probability $P(N)$ can be evaluated to a temperature dependend form factor
\begin{equation}
F(T,\omega) = [1 - \exp\{-(k_{\rm B} T)^{-1} \hbar \omega\}]^{-1} . \label{formfactor}
\end{equation}
The electron-phonon interaction Hamiltonian is
\begin{eqnarray}
\hat V_{\rm PH} & = & \int\! {\rm d}^3q \sum_{\nu a b \sigma} \kappa_{\nu}({\bm q}) \langle a | \exp(-i {\bm q} \hat {\bm r}) | b \rangle \hat \varrho^+_{{\bm q} \nu} \hat a^+_{a \sigma}  \hat a^{~}_{b \sigma} \nonumber\\ & & +\, {\rm h.c.}\, ,
\end{eqnarray}
where $\hat \varrho^+_{{\bm q} \nu}$ is the creation operator of a phonon with wavevector $\bm q$ and branch $\nu$ (longitudinal (l) or transversal (t)). For GaAs, there are two important coupling mechanisms to acoustic phonons: piezoelectric (PE) and deformation potential (DP) coupling. While the former couples to longitudinal and transversal branches the latter at sufficiently low temperatures only couples to longitudinal branches.\cite{ziman} The coupling coefficients therefore are
\begin{eqnarray}
\kappa_{\rm l}({\bm q}) & = & C^{\rm PE}_{\rm l} + C^{\rm DP}_{\rm l} \\
\kappa_{\rm t}({\bm q}) & = & C^{\rm PE}_{\rm t} .
\end{eqnarray}
The DP coupling coefficient is given by\cite{toyabe}
\begin{equation}
C^{\rm DP}_{\rm l} = {( 2 \pi )}^{3/2} D \sqrt{\frac{\hbar}{2 \rho c_l}} \, {|{\bm q}|}^{1/2} .\label{CDP}
\end{equation}
The deformation potential for bulk GaAs is $D = 6.7{\rm eV}$.\cite{adachi} The mass density is $\rho_{\rm GaAs} = 5.65{\rm g}\,{\rm cm}^{-3}$.\cite{rode} The speed of sound for acoustical phonons for GaAs is $c_{\rm l} = 5.11 \cdot 10^5{\rm cm}\,{\rm s}^{-1}$ and $c_{\rm t} = 3.34 \cdot 10^5{\rm cm}\,{\rm s}^{-1}$.\cite{krummheurer} The PE scattering depends inversely on the modulus of the phonon wavevector and therefore is dominant for small phonon energies:\cite{zook,price}
\begin{equation}
C_{\nu}^{\rm PE} = {( 2 \pi )}^{3/2} e\, h_{14} \sqrt{\frac{\hbar}{2 \rho c_{\nu}} A_{\nu}(\theta,\phi)} \, {|{\bm q}|}^{-1/2} ,\label{CPE}
\end{equation}
with the PE tensor element $h_{14}=1.2 \cdot 10^7{\rm V}\,{\rm cm}^{-1}$.\cite{price}
The PE scattering depends strongly on the direction of the wavevector ${\bm q}$ relative to the coordinate system (expressed by the azimuthal angle $\phi$ and the polar angle $\theta$, which is measured with respect to the layer normal). This anisotropy stems from the transformation of the PE tensor from the normal mode to the crystal coordinate system, resulting in the anisotropy functions\cite{zook}
\begin{eqnarray}
A_{\rm l} & = & 36 \alpha^2 \beta^2 \gamma^2 \\
A_{\rm t} & = & 2 \left( \alpha^2 \beta^2 + \beta^2 \gamma^2 + \alpha^2 \gamma^2 \right) - \frac{1}{2} A_{\rm l}(\theta,\phi) .
\end{eqnarray}
$\alpha = \cos(\phi) \sin(\theta)$, $\beta = \sin(\phi) \sin(\theta)$, and $\gamma = \cos(\theta)$ are the direction cosines of the wavevector.

The energy to be absorbed by the phonon bath determines the length of the wavevector for the longitudinal and transversal branches. The possible directions of the wavevector have to be integrated out. The energy $E$ is absorbed by a phonon of wavevector $q = E \hbar^{-1} c_{\nu}^{-1}$. The corresponding phonon density of states is proportional to $q^2$ due to the surface of the sphere in $q$-space, that corresponds to the possible directions of the phonon emission. Eq.~(\ref{phondoppel}) further scales with a factor of $q$ caused by the deformation potential coupling (\ref{CDP}). For very small wavevectors $q$, the piezelectric coupling (\ref{CPE}) is dominant. Therefore, the electron-phonon contribution to the transition rate is proportional to $E$ for small energies and to $E^3$ for larger energies (for very small energies of the order of $k_{\rm B}T$, the form factor Eq.~(\ref{formfactor}) has to be taken into account).

\subsection{The electronic system}

Considering a magnetic field perpendicular to the layer plane, each electronic orbital is a product of a two-dimensional layer wavefunction and a one-dimensional wavefunction for the layer normal direction. For the calculation of the hyperfine and electron-phonon interaction transition matrix elements, the electrons are assumed to be confined in the layer normal direction by an infinitely high potential well. This confinement is considered that much stronger than in the layer plane, that the groundstate of this direction can be assumed to be occupied only.

\begin{figure}
~\\
\includegraphics[width=0.4\textwidth]{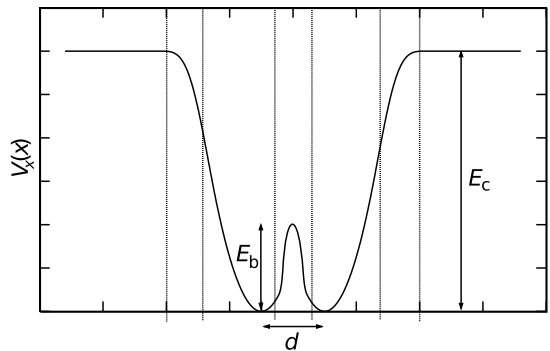}
\caption{\label{potx}Sketch of a double dot confining potential in $x$-direction. The dotted lines define the intervals in which the potential is piecewisely defined (using splines).}
\end{figure}

The Hamiltion describing two-dimensional single-electron orbitals reads:
\begin{equation}
\hat H^{(1)}_{\rm 2D} = \frac{1}{2 m^*} {\left( {\bm p}^{(1)} + e {\bm A}^{(1)} \right)}^2 + V(x,y) ,
\end{equation} 
with the effective electron mass $m^*$ ($6.1 \cdot 10^{-32} {\rm kg}$ for GaAs \cite{landb}), and the vector potential ${\bm A}$ $\,({\bm B} = {\bm \nabla} \times {\bm A})$. The two-dimensional layer orbitals are calculated in a discretized spatial basis, allowing for piecewisely defined (asymmetric) double dot potentials (see Fig.~\ref{potx}). In this way, the curvatures in the dot centers, their distance $d$, and the barrier height $E_{\rm b}$ can be defined independently. We define the dots to be parabolic near the dot centers. The confinement in $y$-direction is assumed to be parabolic.

Outside the dot regions we consider the confining potential to be constant $V(x,y) = E_{c}$. We choose this constant so large that no unbound states are mixed by the Coulomb interaction to the low-energetic orbitals we consider. In this region we choose a sparser grid than in the dot region, because the wave functions decay slowly here.

The Coulomb interaction between the electrons is calculated in the layer plane only, again requiring a strong confinement perpendicular to the layer. The corresponding matrix elements are calculated in momentum space, taking advantage of the inverse distance operator being diagonal in momentum space:
\begin{equation}
\frac{1}{2 \pi} \int\!{\rm d}^2r\, \frac{1}{r} \exp(-i{\bm q}{\bm r}) = \frac{1}{q} .
\end{equation}
Hereby, only two-dimensional integrals have to be evaluated instead of four-dimensional integrals in real space.

Since the Coulomb interaction does not affect the electron spin, the resulting two-electron Hamiltonian can be diagonalized in blocks of fixed pairs ($S^2,S_z$). The few-electron states are calculated by exact diagonalization in a Fock basis. `Exact' means here, that all correlations due to the Coulomb interaction within the finite basis set are taken into account. The Zeeman energy corresponding to the two-electron spin component $S_z$ simply adds to the orbital energies.

We have considered single- and two-electron systems in single and double dots. The systems have a confinement of $\hbar \omega_0 \approx 2 {\rm meV}$. The double dots have a distance of about $20 {\rm nm}$ between the dot centers, separated by an energy barrier of $20 \rm{meV}$ (due to the energy barrier between the dots the double dot potential is deformed, such that the single-particle ground state is increased to $\hbar \omega_0 \approx 2.75 {\rm meV}$). For the layer thickness of the dot system we have chosen a typical value of $2.8 {\rm nm}$, corresponding to ten monolayers of GaAs.\cite{wu} All calculations have been performed for the temperature $T = 0.1 {\rm K}$.

\section{Numerical results}

\subsection{Transitions in single-electron systems}

\begin{figure}
\vspace{1.1em}
a)\hfill{}~\\
\vspace{-3em}\includegraphics[angle=-90,scale=0.34]{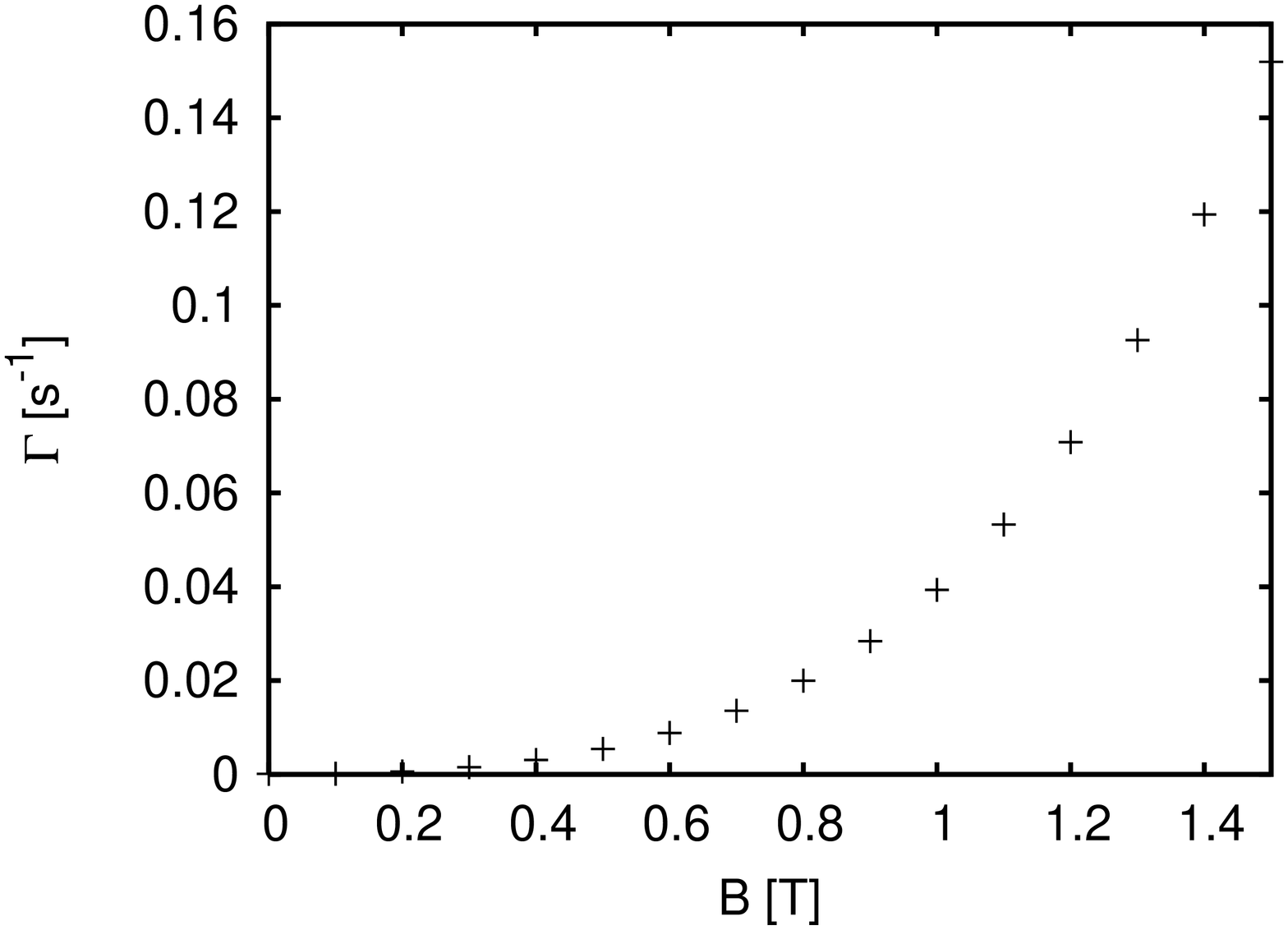}
~\\ ~\\
b)\hfill{}~\\
\vspace{-3em}\includegraphics[angle=-90,scale=0.34]{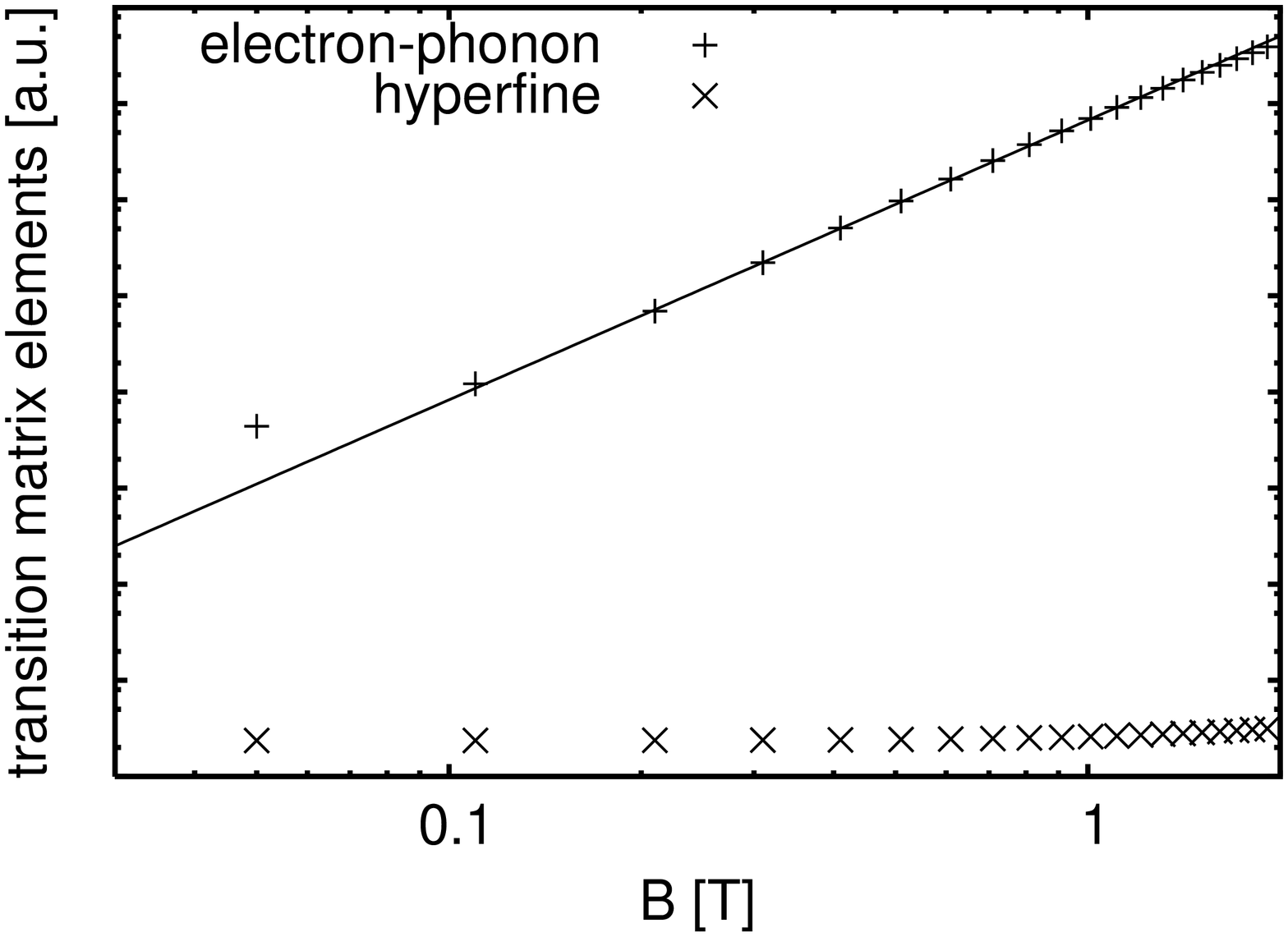}
\caption{\label{rate1in1sz}Second order transition rate (a) from the spin down state to the spin up state of an electron occupying the lowest orbital of a GaAs quantum dot and logarthmic plots of electron-phonon and hyperfine transition matrix elements (b) (arbitrarily scaled). The fitted curve is a straight line with slope $\approx 3$.}
\end{figure}

For single-electron systems, we consider $S_z$ transitions of an electron in the lowest orbital. For $B=0$ the corresponding two states $| S_z = \pm 1 \rangle$ are degenerate. The second order transition rate is zero here (see Fig.~\ref{rate1in1sz}(a)), because the phonon density of states vanishes for $q \rightarrow 0$. Furthermore, the operator $\exp(i {\bm q} \hat {\bm r})$ approaches $\hat 1$, such that different states are not coupled. At $B=0$ a first order, ``hyperfine-only'' transition is possible. For arbitrarily small magnetic fields $B>0$, the electronic Zeeman energy is always larger than the nuclear Zeeman energy, forbidding the first order relaxation due to the law of energy conservation.

The second order transition rate increases with the magnetic field, since the electron-phonon
and hyperfine transition elements increase. In addition, the energies of states with higher angular momentum decrease, leading to smaller energy dominators in Eq.~(\ref{goldenrule}). Therefore, these states contribute more to the transition rate with increasing magnetic field.

Fig.~\ref{rate1in1sz}(b) shows the electron-phonon transition element from the initial state to the state with the same total spin in the first excited orbital. The logarithmic plot reveals that the absolute square of the element is proportional to $B^3$, {\it i.e.}~to the cube of the electronic Zeeman energy. For small $q$, the electron-phonon matrix element is proportional to $B$ due to dominating piezoelectric coupling.

The corresponding hyperfine transition element (Fig.~\ref{rate1in1sz}(b)) leads the system back from the
intermediate state, which we have considered above for the electron-phonon interaction, back to the final state. The hyperfine transition elements grow mostly due to the fact that the magnetic field shrinks the electronic wavefunctions. The matrix elements are proportional to an integral over the product of four electronic wavefunctions (cf.~Eq.~(\ref{vhffourorb})). For wavefunctions of a harmonic oscillator, such an integral is inversely proportional to the width of the wavefunctions. Since the width is reduced with increasing magnetic field, this explains the growth of the hyperfine transition matrix elements.

The second order transition rate of a single electron in a {\em double dot} is similar to the single-dot case. We have found the rate to be two orders of magnitude larger than for the single dot. This is due to the energetic differences to the next intermediate states. In the double-dot case, this energy is smaller due to tunnel splitting, resulting in a larger transition rate.

\subsection{Transitions in two-electron systems}

Spin transitions in two-electron systems can occur within the triplet sector, {\it i.e.}\ between triplets with $S_z \in \{0,1\}$, or between triplet and singlet states. For a two (electron) spin system, $S_z$ transitions in the triplet sector are decoupled from the orbital degrees of freedom.

\subsubsection{$S_z$ transitions}

\begin{figure}
\includegraphics[angle=-90,scale=0.34]{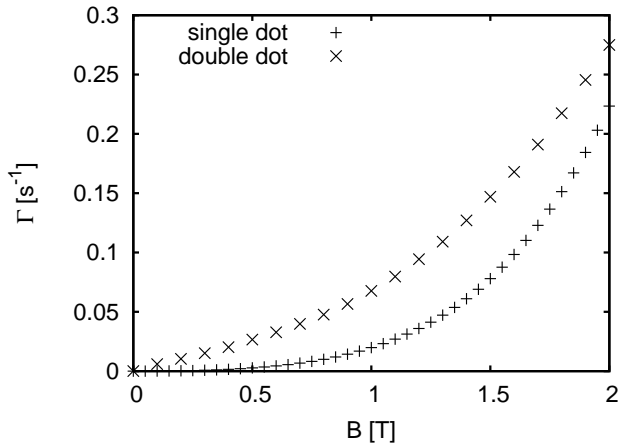}
\caption{\label{rate2inxsz}Second order transition rates between the lowest energetic two-electron triplets with $S_z=1$ and $S_z=0$ in a single and a double GaAs quantum dot, respectively.}
\end{figure}

Fig.~\ref{rate2inxsz} shows the second order rates between the lowest energetic two-electron triplets with $S_z=1$ and $S_z=0$ for a single dot and a double dot, respectively. Since the two triplets are degenerate at $B=0$, the second order rate is zero there. Again, a first order transition is possible here, of course, which is not considered here. A transition within the triplet sector is a simple spinflip. As in the single-electron case it does not involve any orbital transitions. Therefore the rate has the same magnetic field dependence.

\subsubsection{$S^2$ transitions in single dots}

In contrast to the spin flip processes considered so far, the singlet-triplet transitions involve the orbital parts of the wave functions.

\begin{figure}
\vspace{1.1em}
a)\hfill{}~\\
\vspace{-3em}\includegraphics[angle=-90,scale=0.34]{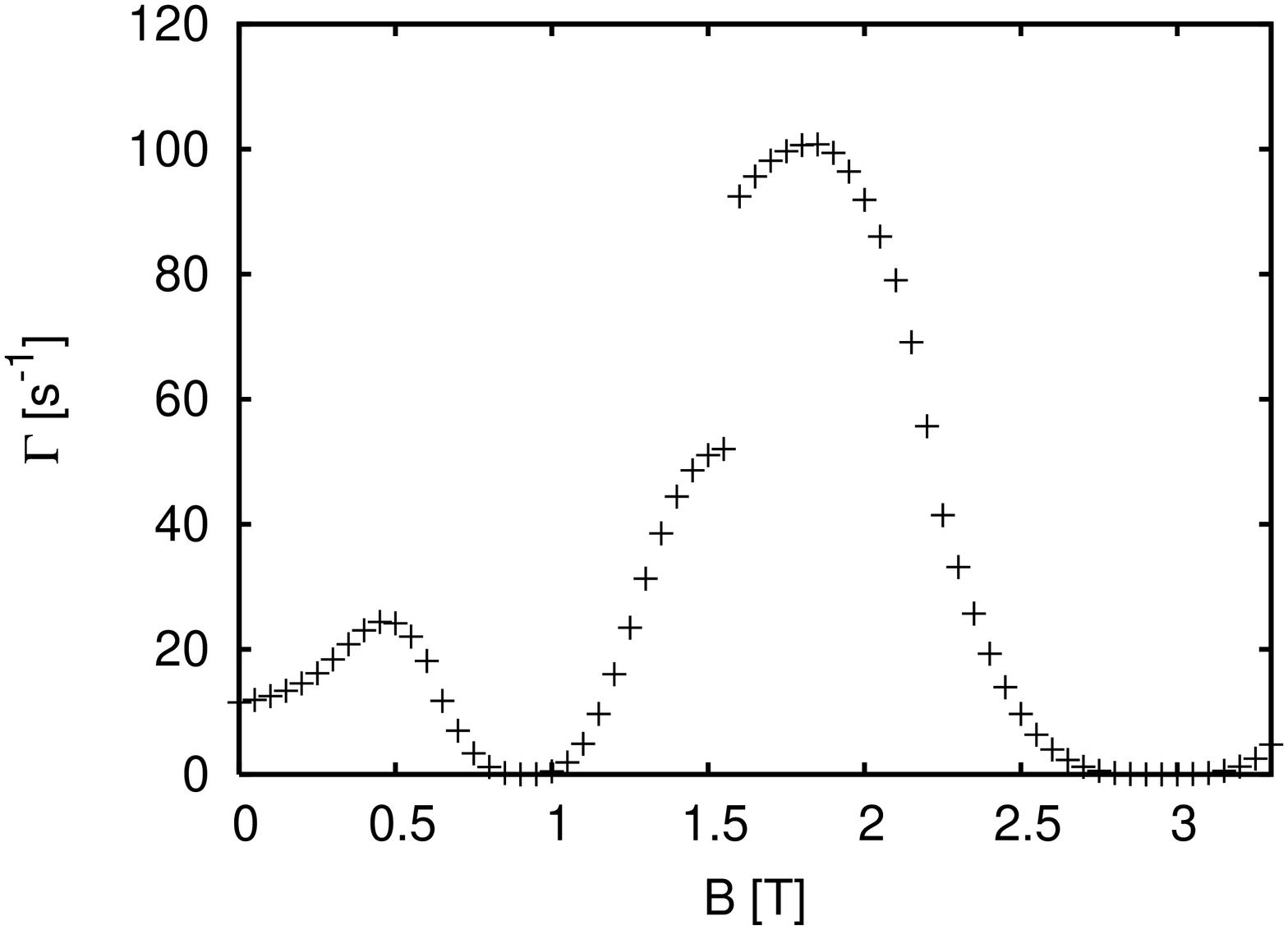}\\
~\\ ~\\
b)\hfill{}~\\
\vspace{-3em}\includegraphics[angle=-90,scale=0.34]{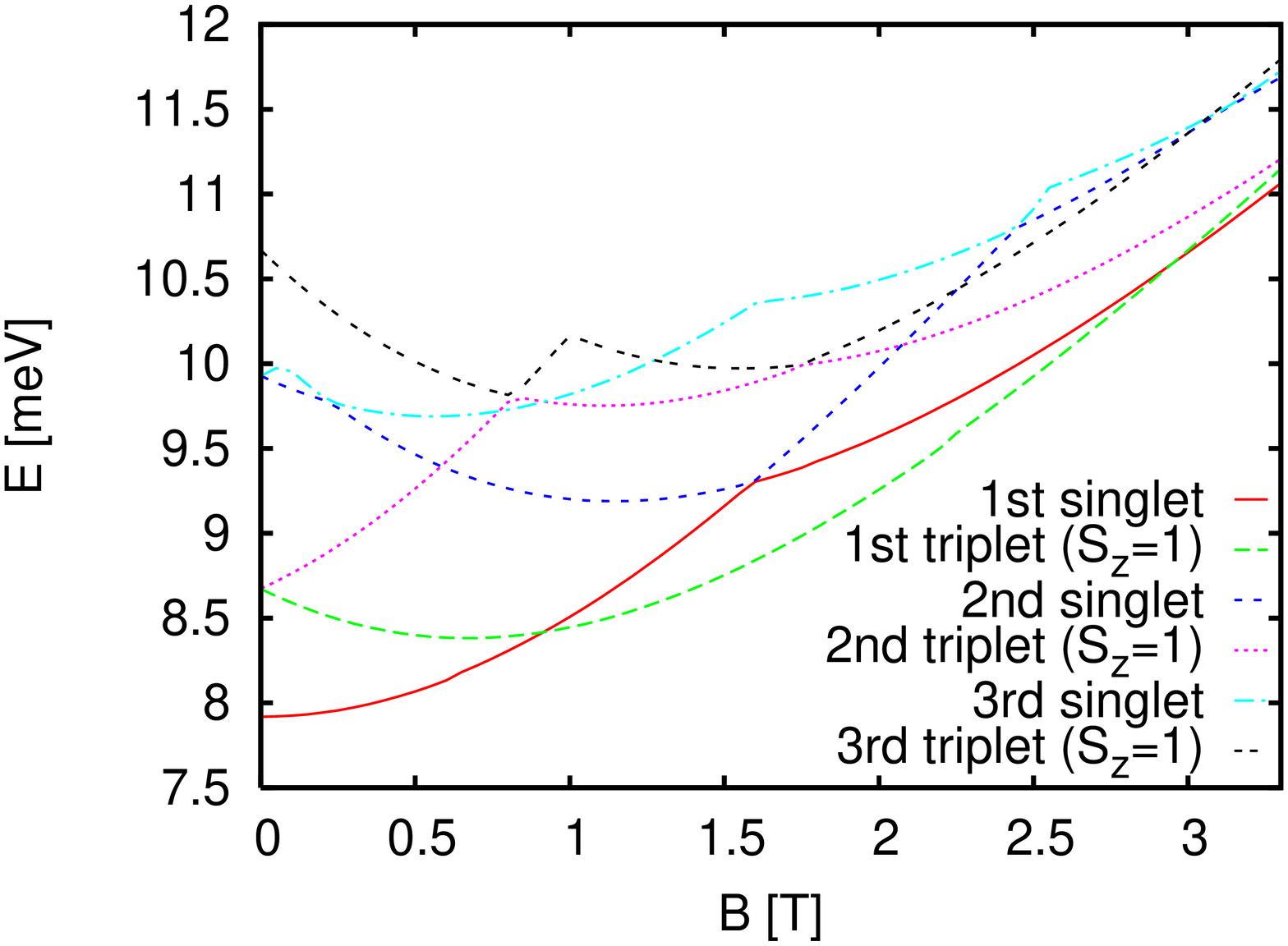}
\caption{\label{rate2in1s2nc}(Color online) Second order transition rate (a) between the lowest energetic two-electron singlet and triplet with $S_z=1$ in a GaAs quantum dot (normal Coulomb interaction: $\epsilon = 13.1$) and energy spectrum (b) (only the lowest energies are shown; for the rate calculations a few hundreds of intermediate states have to be included for convergence of the rates).}
\end{figure}

\begin{figure}
\vspace{1.1em}
a)\hfill{}~\\
\vspace{-3em}\includegraphics[angle=-90,scale=0.34]{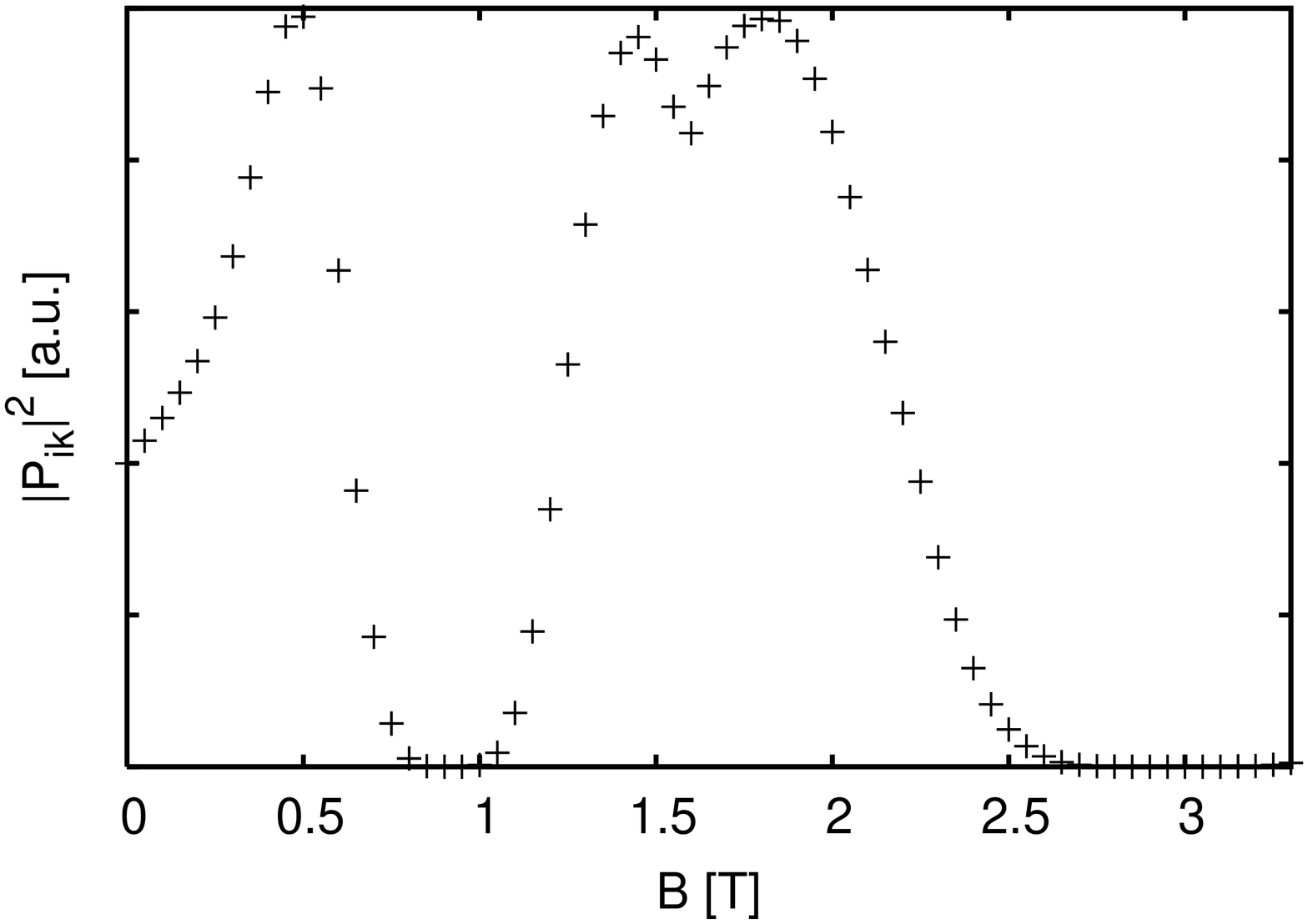}
~\\ ~\\
b)\hfill{}~\\
\vspace{-3em}\includegraphics[angle=-90,scale=0.34]{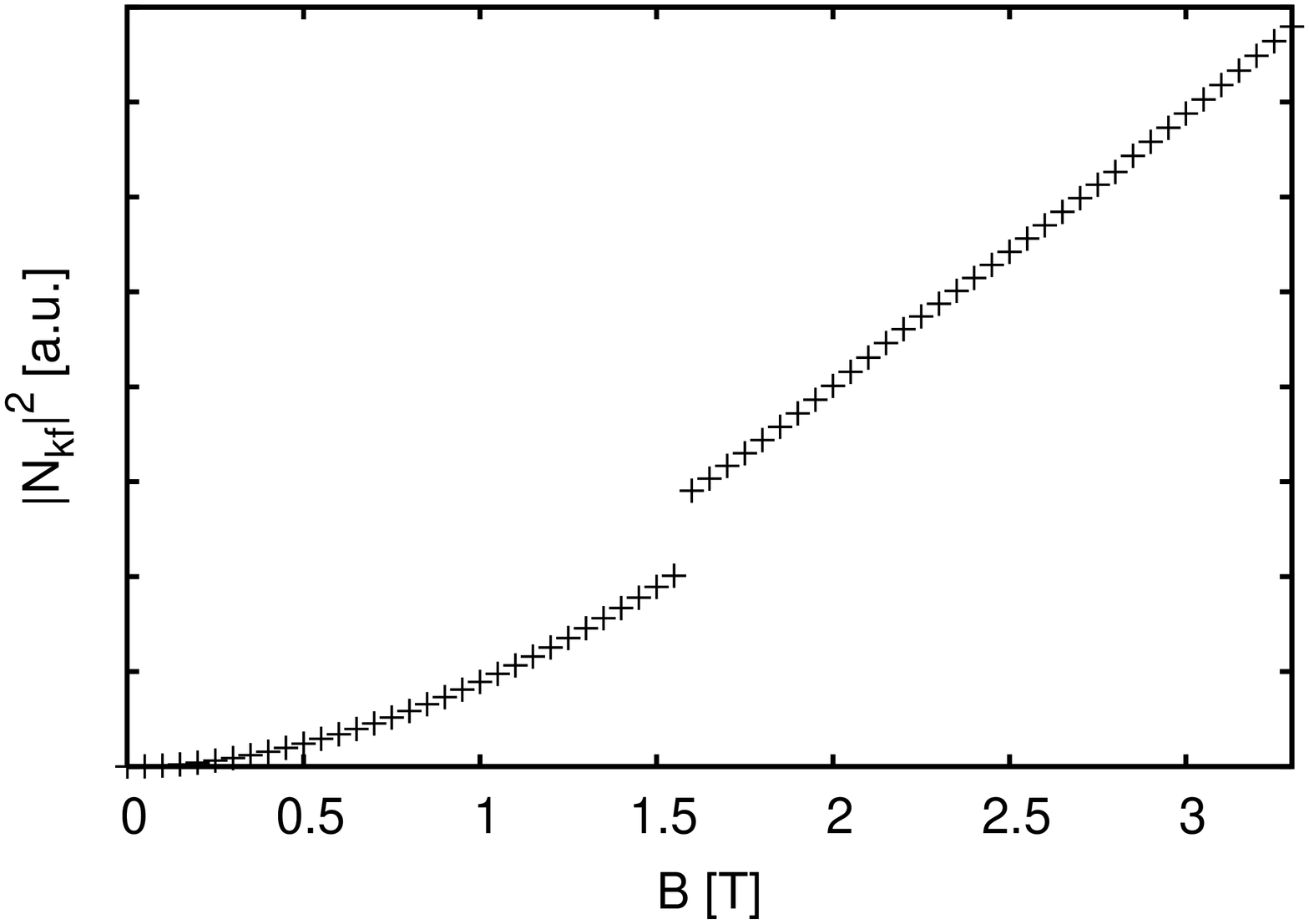}
\caption{\label{phon2in1s2nc}Modulus square of the electron-phonon transition matrix element (a) from the initial electronic state to the state in the next orbital with equal spin (two electrons in a GaAs quantum dot; normal Coulomb interaction) and modulus square of the hyperfine transition matrix element (b) from the intermediate state to the final state.}
\end{figure}

Fig.~\ref{rate2in1s2nc} shows the second order $S_z${}$=$1-triplet-singlet transition rate (a) and the energy spectrum (b) (cf.~[\onlinecite{ellenberger:126806,PhysRevB.66.035320}]) of two electrons in a GaAs quantum dot with ``normal'' Coulomb interaction ($\epsilon=13.1$; in order to study the influence of the interaction strength, we will rescale this value). At $B=0$ the second order rate is finite, since the energies of the singlet and the triplet differ and thus, a nonzero phonon density of states enters the rate. In the range of the magnetic field we studied, the rate shows two minima and two maxima. This behavior of the rate is caused by the electron-phonon interaction, as the transition matrix elements have minima and maxima at the corresponding magnetic fields (see Fig.~\ref{phon2in1s2nc}(a)). Though the triplet-singlet energy difference decreases for small, increasing magnetic fields, the electron-phonon transition matrix element grows with the phonon density of states for magnetic fields up to $0.5 {\rm T}$. The reason for this is, that with the decreasing energy to be dissipated and the corresponding phonon wavevector of decreasing length, a geometric resonance with the dot size is approached. Hence, the corresponding single-particle transition matrix elements grow. For higher magnetic fields, approaching the triplet-singlet crossing, the vanishing phonon density of states dominates the transition rate. In the vicinity of the triplet-singlet crossings shown here, at $B=0.92 {\rm T}$ and $B=2.95 {\rm T}$, it causes the second order transion rate to vanish (Meunier {\it et al.}\cite{meunier:126601}\ have experimentally observed triplet-singlet relaxation caused by electron-phonon interaction with an enhanced spin lifetime near triplet-singlet crossings). Between the crossings, the rate follows the energy dependence of the gap between singlet and triplet. In the vicinity of the triplet-singlet crossing, resonances in the transition rate can be expected due to first order, ``hyperfine-only'' relaxation.\footnote{Inoshita, Ono, and Tarucha \cite{inoshita} have calculated the evolution of a hyperfine interacting electron spin in Born-Markov approximation.\cite{blum} They have found resonances in the transverse relaxation $1/T_2$ for triplet-singlet crossings.}

Due to the increasing {\em hyperfine transition} elements (see Fig.~\ref{phon2in1s2nc}(b)) the second maximum of the second order transition rate is higher than the first maximum. The electron-phonon coupling thus mainly is responsible for the vanishing second order rate near the energy crossings, while the hyperfine interaction is responsible for the overall growth. The explanation for the growth of the hyperfine transition matrix elements is again the reduction of the width of the electronic wavefunctions with increasing magnetic field.

At about $1.6 {\rm T}$ a crossing between the initial singlet state and the next singlet occurs. This leads to a pronounced effect on the transition rate: the rate ``jumps'' by about $75\%$. This is due to the drastic change of the character of the initial state. At this magnetic field the excited singlet state, from which the spin transition to the ground state is considered, changes. While the change of the electron-phonon transition elements at the corresponding magnetic field is not as dramatic (see Fig.~\ref{phon2in1s2nc}(a)), the hyperfine transition elements show a ``jump'' there (see Fig.~\ref{phon2in1s2nc}(b)). This is amplified by a large electron-phonon rate which changes continuously at this transition. The crossing in the initial state does not lead to a resonance, caused by a vanishing energy denominator in the transition rate formula, as this intermediate state is not an intermediate state that allows for a second order transition to the triplet ground state.

We have studied the behavior of the second order transition rate for systems differing in the strength of the Coulomb interaction (modeled by a variation of the dielectric constant $\epsilon$).
The second order rate is minimal for the case of ``normal'' Coulomb interaction ($\epsilon = 13.1$). This is in clear distinction of what one would expect for first order rates. In the first order rates, only the hyperfine transition matrix elements enter. These are lowered due to correlations between the electrons caused by a stronger Coulomb interaction ($\epsilon = 6.55$). In the second order rate the correlation induced reduction of the hyperfine rate is compensated by the increasing electron-phonon matrix elements, which grow with the triplet-singlet energy difference. Therefore, an increase of the Coulomb energy need not necessarily lead to an increased lifetime of the electron spin, if the dominant relaxation mechanism is the combined hyperfine and electron-phonon scattering.

\subsubsection{$S^2$ transitions in symmetric double dots}

\begin{figure}
\vspace{1.1em}
a)\hfill{}~\\
\vspace{-3em}\includegraphics[angle=-90,scale=0.34]{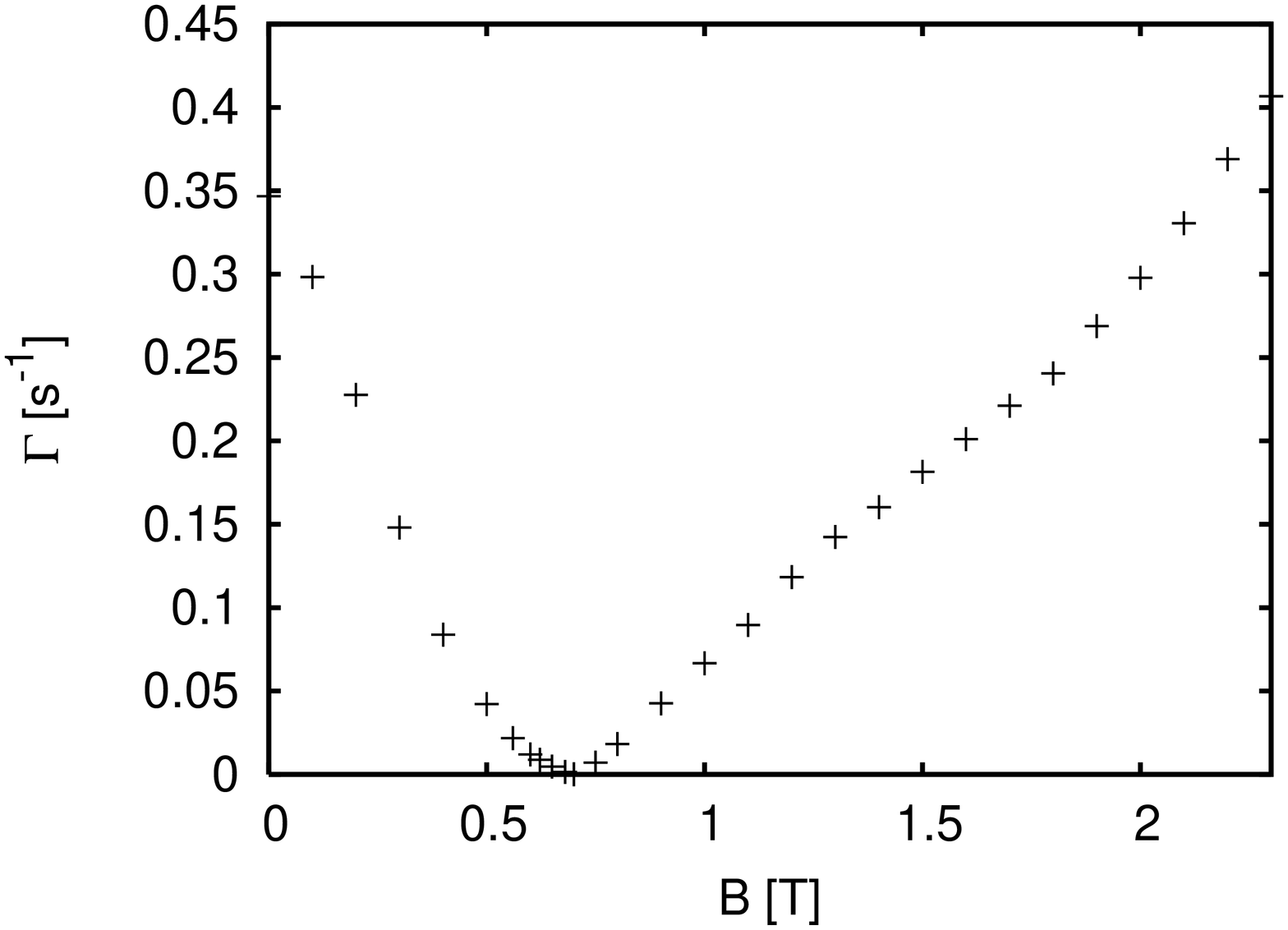}
~\\ ~\\
b)\hfill{}~\\
\vspace{-3em}\includegraphics[angle=-90,scale=0.34]{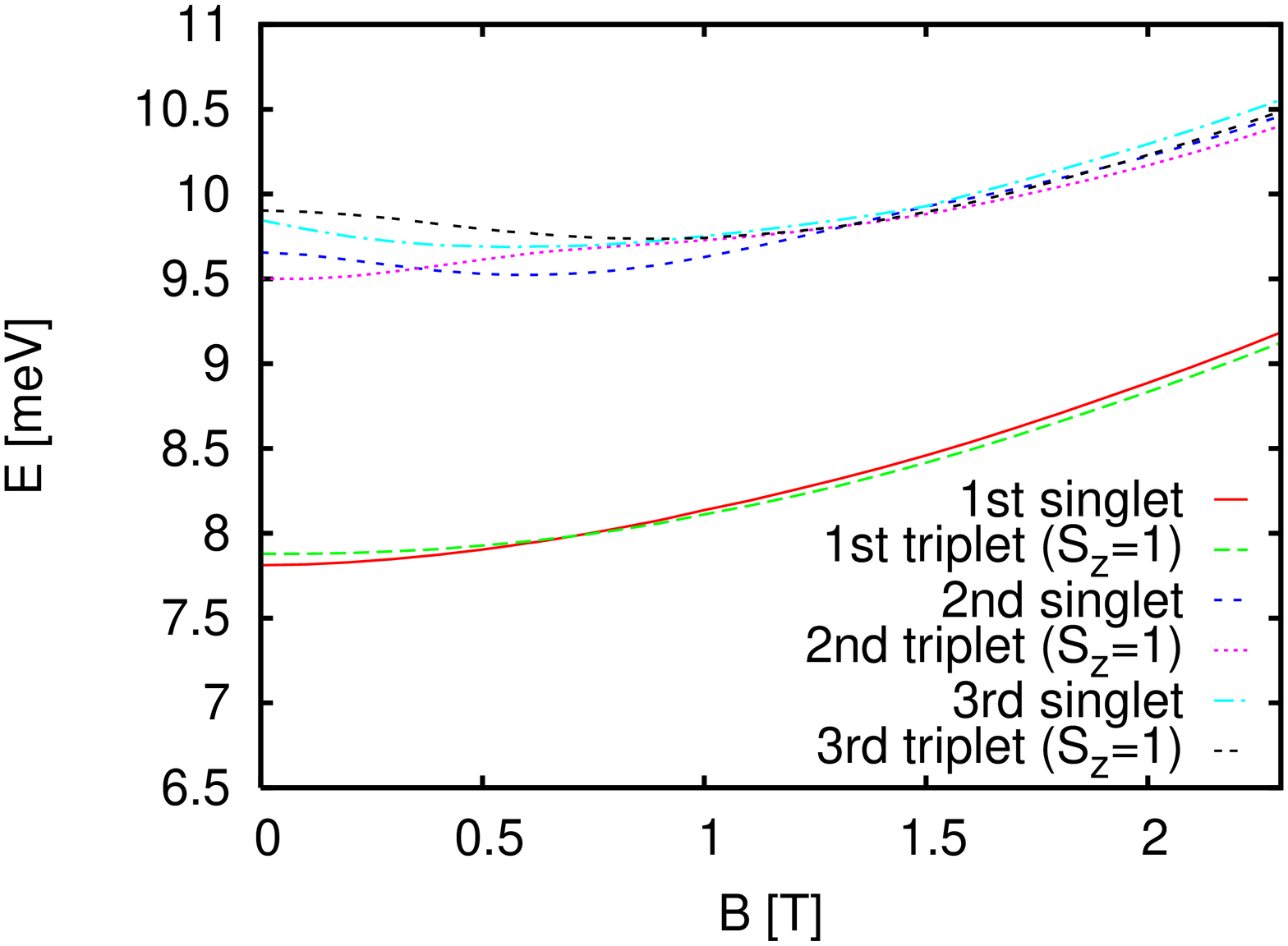}
\caption{\label{rate2in2s2nc}(Color online) Second order transition rate (a) between the lowest energetic two-electron singlet and triplet with $S_z=1$ in a GaAs double quantum dot (normal Coulomb interaction: $\epsilon = 13.1$) and energy spectrum (b).}
\end{figure}

Fig.~\ref{rate2in2s2nc} shows the second order triplet-singlet transition rate (a) and the energy spectrum (b) of two electrons in a GaAs double quantum dot with ``normal'' Coulomb interaction. Since the lowest energetic singlet and triplets are separated from the next singlet and triplets, crossings of intermediate states with the initial state, if any, could only occur at very high magnetic fields, which we have not considered here. The ``jump'' we have found in the single-dot rates, will thus not appear in the following plots.

As for the single dot, the triplet and singlet states are non-degenerate and thus the second order rate is finite for $B=0$. Due to the Coulomb interaction the dots are essentially singly occupied. Thus the energy difference between singlet and triplet ground state is small, and thus, no geometric resonance can occur in the electron-phonon transition matrix elements. Therefore, the transition rate monotonously falls approaching the triplet-singlet crossing. Expectedly, the second order rate vanishes in the vicinity of the triplet-singlet crossing. Due to a nearly constant energy gap, the overall growth of the rate beyond the triplet-singlet crossing is due to the increasing hyperfine transition matrix elements.

Owing to the tunnel splitting, the energy differences between the initial and final states in the double-dot case are smaller than in the single-dot case. As this causes a smaller phonon density of states entering the second order transition rate, the rate for the double dot system is smaller than the transition rate for the single-dot system. For $B=0$, {\it e.g.}, the transition rate in the single dot-system is about one order of magnitude larger. Furthermore, the next singlet and triplet states are separated from the lowest ones, which means a lower contribution of these states to the second order rate.

\begin{figure}
\vspace{1.1em}
a)\hfill{}~\\
\vspace{-3em}\includegraphics[angle=-90,scale=0.34]{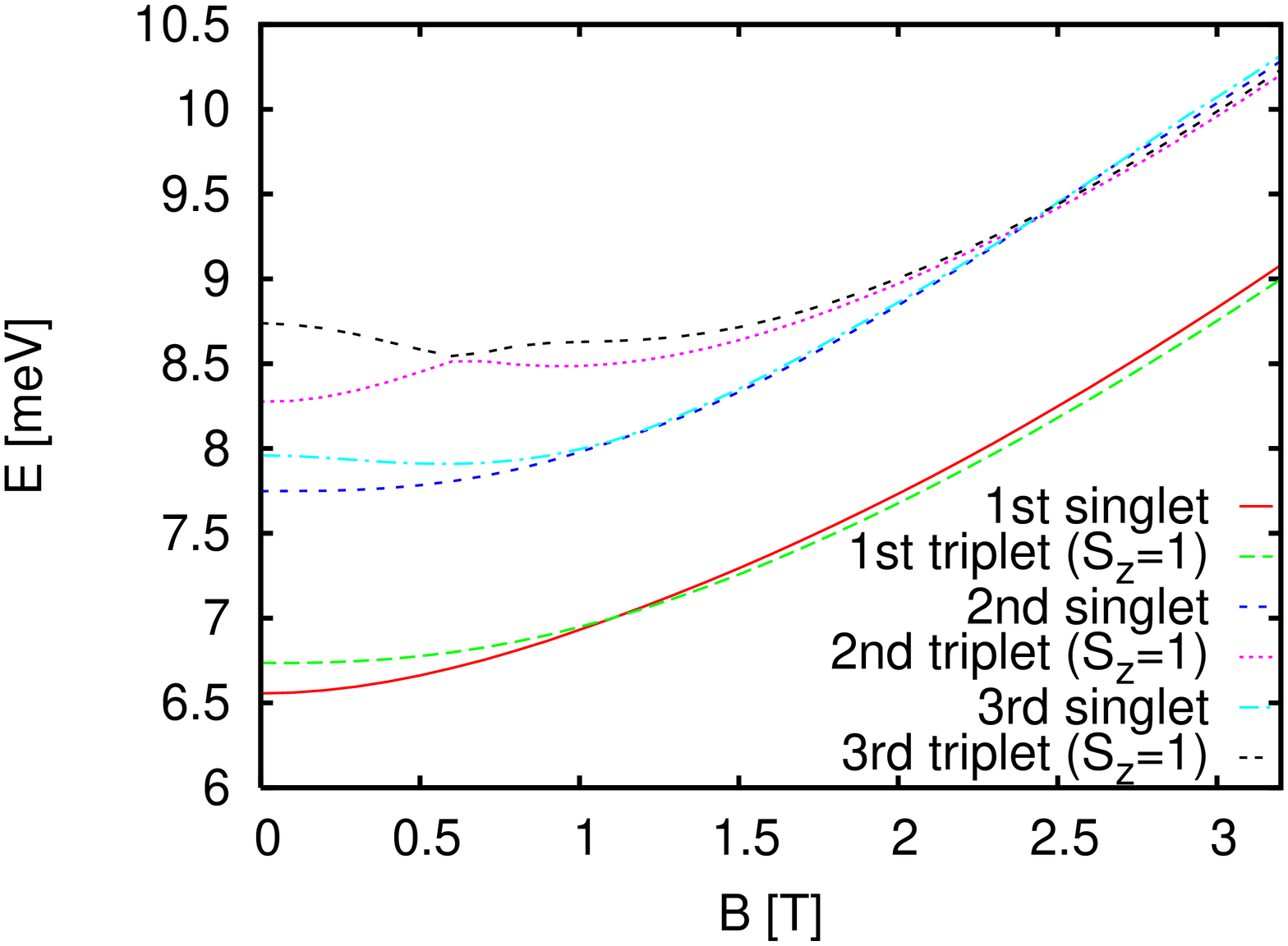}
~\\ ~\\
b)\hfill{}~\\
\vspace{-3em}\includegraphics[angle=-90,scale=0.34]{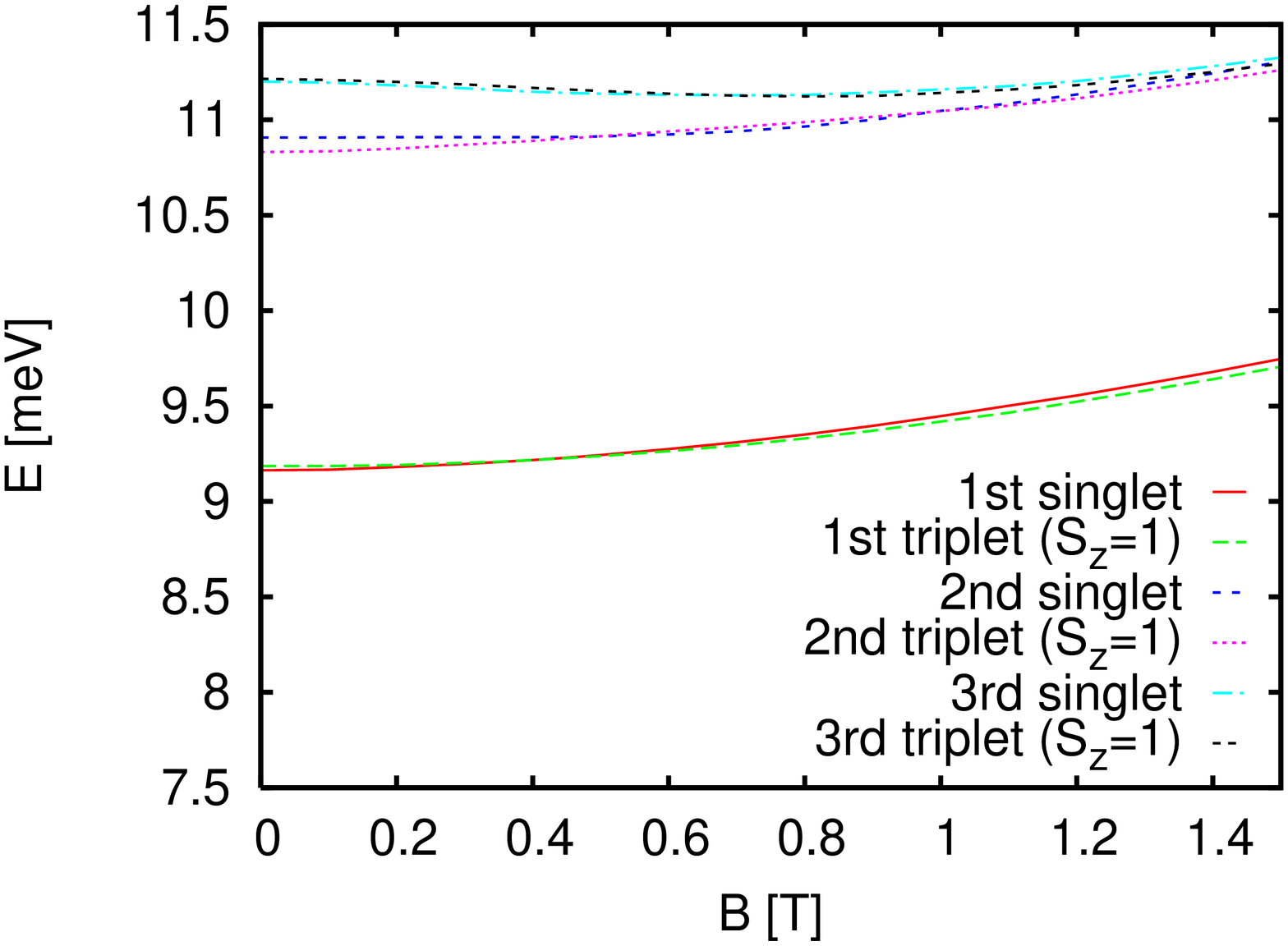}
\caption{\label{rate2in2s2lc}(Color online) Energy spectrum the lowest energetic two-electron singlets and triplets with $S_z=1$ in a GaAs double quantum dot (weak Coulomb interaction: $\epsilon = 39.3$ (a) and strong Coulomb interaction: $\epsilon = 6.55$ (b)).}
\end{figure}

For weak Coulomb interaction ($\epsilon=39.3$) the second order transition rate is similar to the case of normal Coulomb interaction. However, due to differences in the spectrum (see Fig.~\ref{rate2in2s2lc}(a)) the rates at low magnetic fields are $\sim$10 times larger than for normal Coulomb interaction.

For strong Coulomb interaction ($\epsilon=6.55$) the second order transition rate at small magnetic fields is about $10$ times smaller than the rate in the case of normal Coulomb interaction, although the energy differences to the intermediate states are of the same magnitude (see Fig.~\ref{rate2in2s2lc}(b)). There are two reasons for this. Firstly, the overall energetic distance between the initial and final state is lower for the case of strong Coulomb interaction. Secondly, correlations in the electronic wavefunctions due to the Coulomb interaction have the effect of a reduction of the transition matrix elements.

\subsubsection{$S^2$ transitions in asymmetric double dots}

\begin{figure}
\vspace{1.1em}
a)\hfill{}~\\
\vspace{-3em}\includegraphics[angle=-90,scale=0.34]{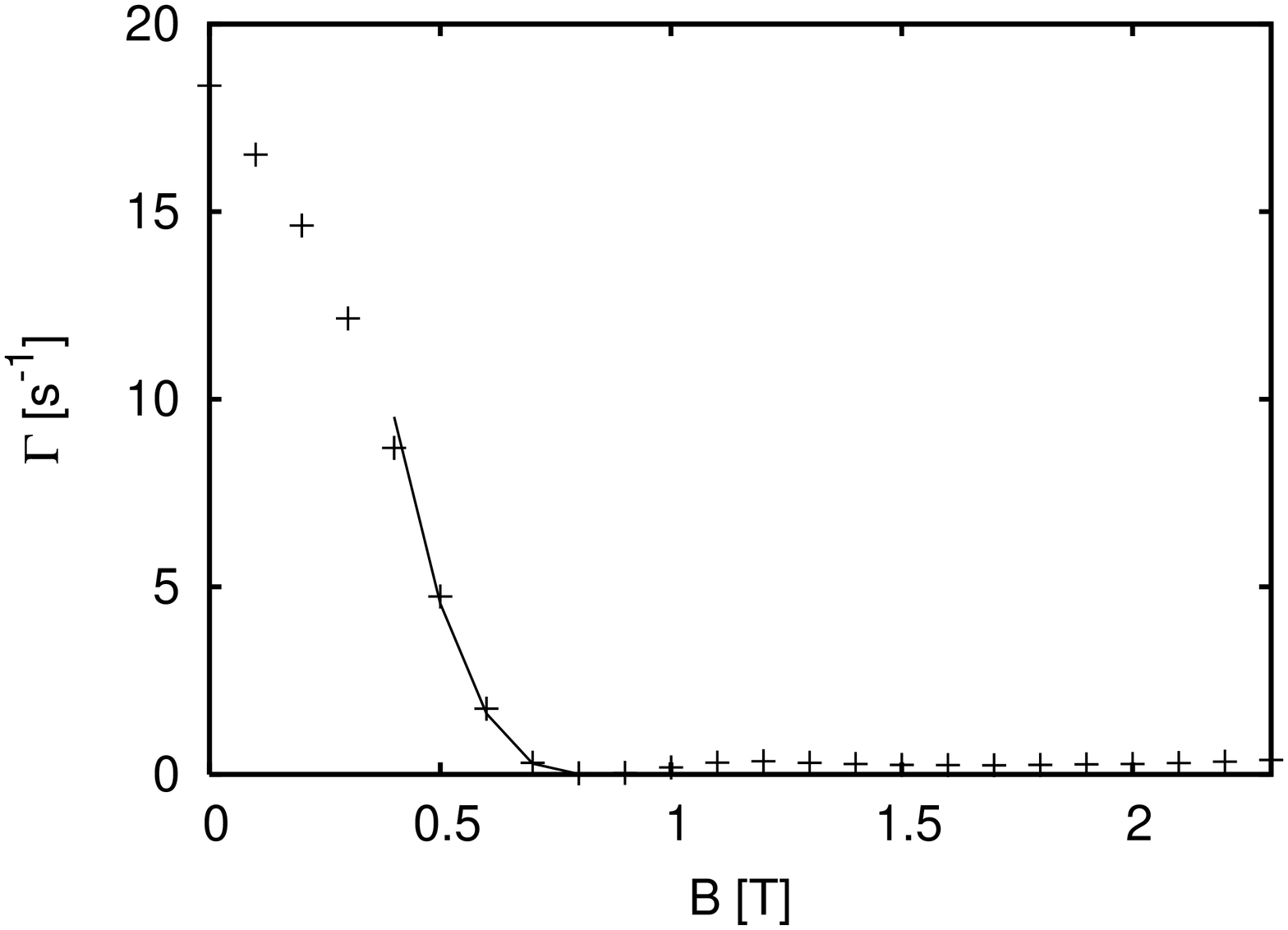}
~\\ ~\\
b)\hfill{}~\\
\vspace{-3em}\includegraphics[angle=-90,scale=0.34]{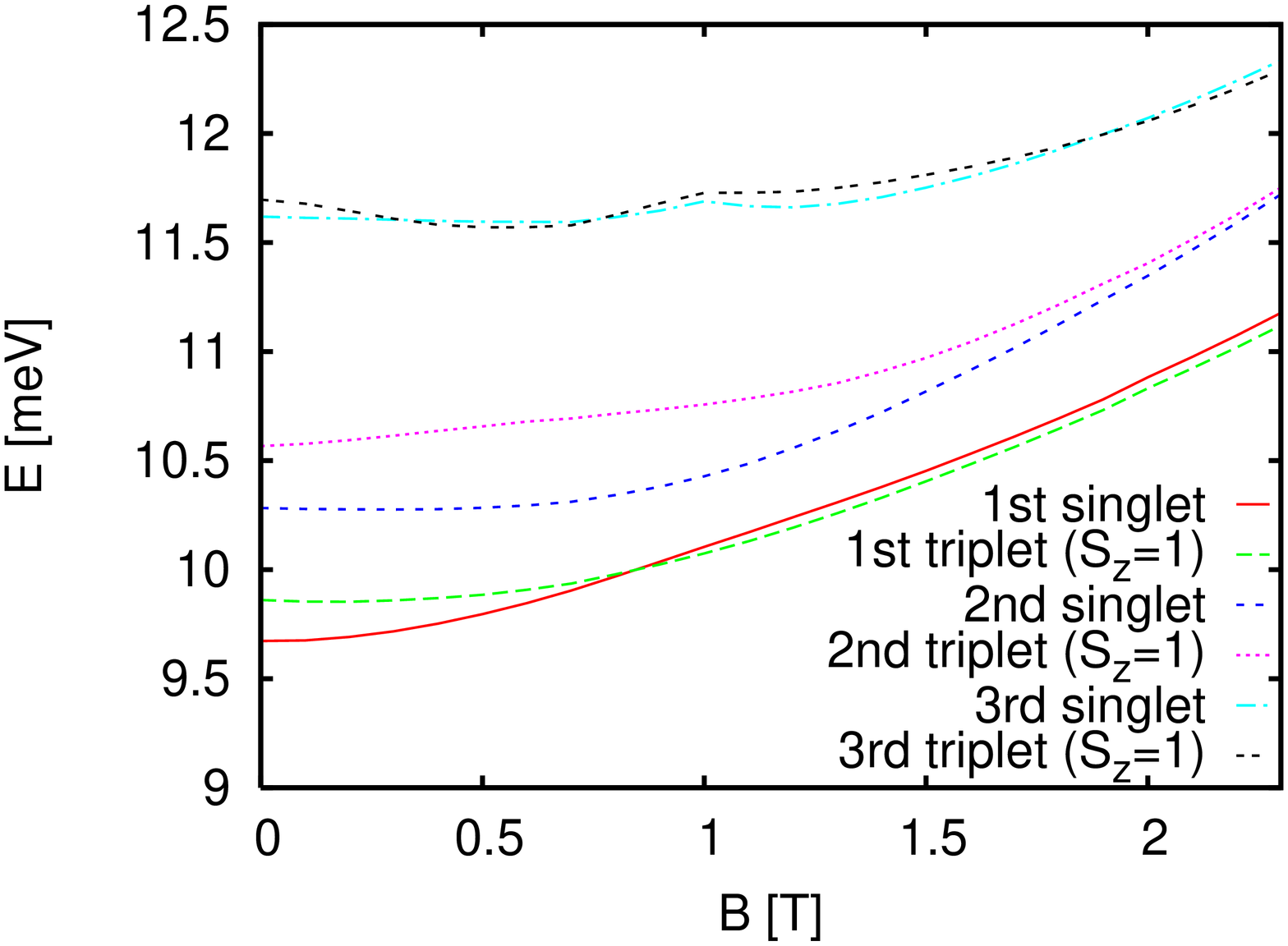}
\caption{\label{rate2in2s2asymm}(Color online) Second order transition rate between the lowest energetic two-electron singlet and triplet with $S_z=1$ in a highly asymmetric GaAs double quantum dot (normal Coulomb interaction: $\epsilon = 13.1$) and energy spectrum.}
\end{figure}

\begin{figure}
\includegraphics[scale=0.36]{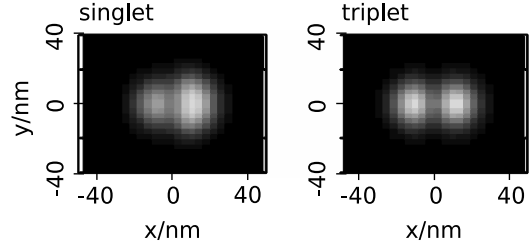}
\caption{\label{denssingtrip}Electron densities of groundstate singlet and lowest energetic triplet with $S_z=1$ for the highly asymmetric double quantum dot at $B = 0$.}
\end{figure}

Fig.~\ref{rate2in2s2asymm} shows the second order two-electron triplet-singlet transition rate (a) and the energy spectrum (b) for a double dot with the right dot being energetically $2 {\rm meV}$ lower than the other dot. Due to this asymmetry, for $B=0$ (almost) no tunneling is possible between the left and right dot orbitals (we recall that the effective single-dot single-particle energy is about $2.75 {\rm meV}$). This means a larger energy difference between the singlet and the triplet than in the case of a symmetric double dot, and thus, the second order rate is larger.

\begin{figure}
\includegraphics[angle=-90,scale=0.34]{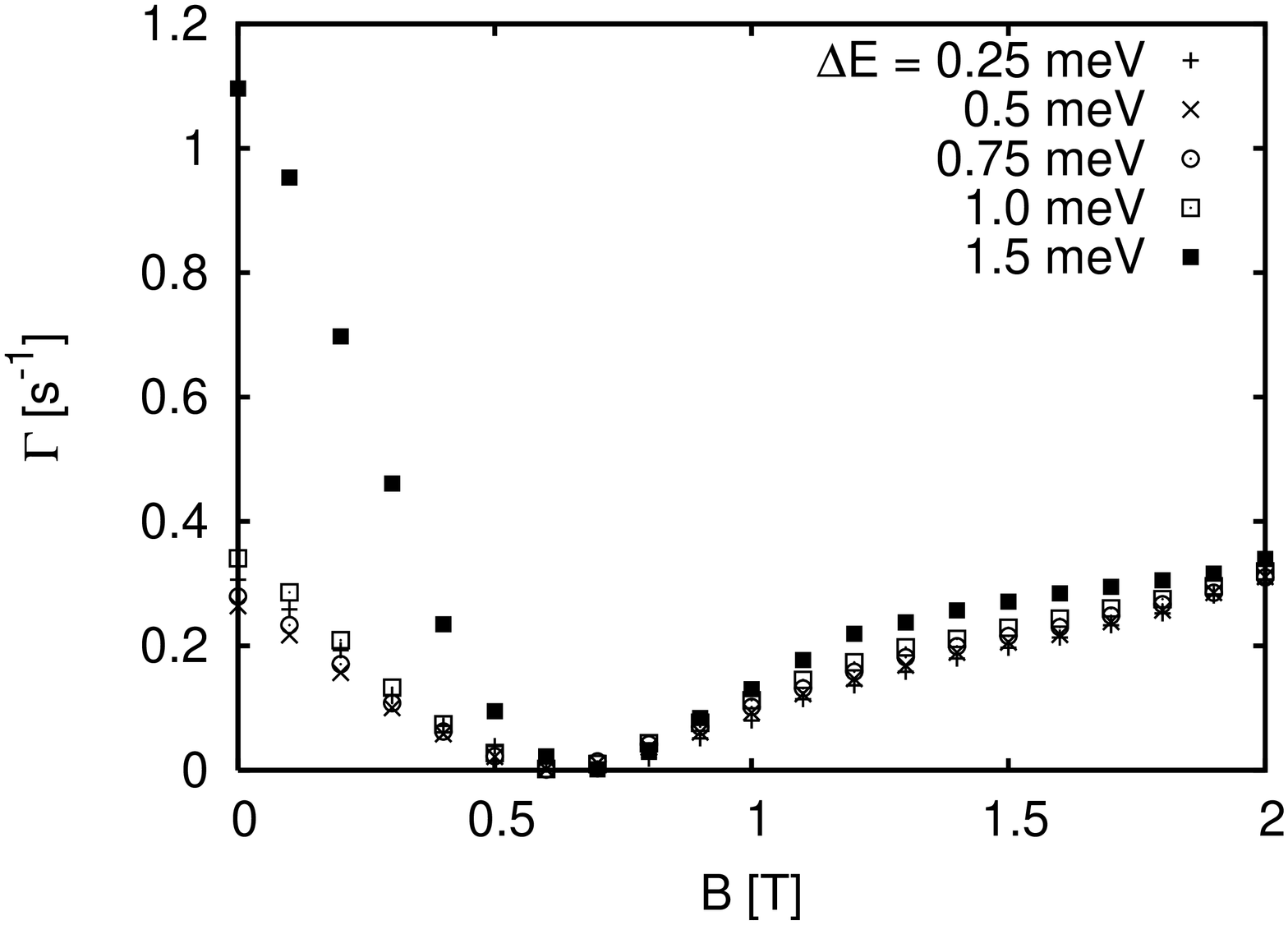}
\caption{\label{ratecamps2asymm}Second order transition rate between the lowest energetic two-electron singlet and triplet with $S_z=1$ for different degrees of asymmetry.}
\end{figure}

For $B=0$ the two-electron ground state is a singlet that has a large electron density in the right dot, but also a non-negligible density in the energetically higher left dot. The lowest triplet shows a relatively symmetric occupation of both dots (see Fig.~\ref{denssingtrip}). This is plausible, since the first excited single-particle orbital of the asymmetric double dot for $B=0$ mostly is given by the left dot ground state. Of course, the Coulomb energy of the triplet ($\langle \hat V_C \rangle \approx 2.0 {\rm meV}$) is much smaller than the Coulomb energy of the singlet ($\langle \hat V_C \rangle \approx 4.5 {\rm meV}$) due to the larger distance between the electrons. At a magnetic field of $B_{\rm cross}=0.84 {\rm T}$ the reduction of the Coulomb energy becomes that advantageous, that a triplet-singlet crossing occurs and the triplet becomes the ground state. With increasing magnetic field, the energy of the first excited single-particle state in the left and the right dot is lowered (angular momentum $m=-1$). When the first excited state in the right dot approaches the ground state of the left dot, tunneling between these two states becomes possible. The tunnel-splitting leads to smaller energy differences between the initial and final electronic states. Therefore, the second order transition rate is reduced and remains relatively small for magnetic fields $B > B_{\rm cross}$.

In Fig.~\ref{ratecamps2asymm} triplet-singlet transition rates for different degrees of asymmetry are compared. For small energy differences between the left and right dot, tunneling is also possible for small magnetics fields $B \rightarrow 0$. This leads to small second order rates. For larger energy differences between the dots, only with increasing magnetic field tunneling becomes possible.

Fitting the data of the rate (see Fig.~\ref{rate2in2s2asymm}) for small magnetic fields $B<B_{\rm cross}$, we find that the rate is proportional to $\Delta_{\rm STr}^{2.7}$, where $\Delta_{\rm STr}$ is the energy difference between the singlet and the triplet. Since the electron-phonon transition matrix elements entering the second order rate approximately depend on the cube of this energy difference, the result of the fit confirms the assumption, that the electron-phonon interaction is responsible for the decrease of the transition rate.

Only in the vicinity of the crossing, the energy difference is small enough to allow for a first order, ``hyperfine-only'' relaxation. Hence, if the magnetic field of the crossing is avoided, the singlet and the triplet can be considered relatively stable within the corresponding range of the magnetic field. This tunable behavior is desirable for the manipulation of quantum gates.

\section{Conclusion}

Our numerical results for electron spin relaxation rates in quantum dots reveal to be remarkably influenced by correlations in the electronic wavefunctions, caused by the Coulomb interaction between the electrons. The second order rates further strongly depend on the electronic energy spectra.

While first order rates can be considered to be lowered due to correlations caused by a stronger Coulomb interaction, the influence on the second order relaxation rates, mediated by hyperfine plus electron-phonon interaction, is more complex due to dependence on energy differences between the electronic states.

For energetic crossings between the initial and final electronic states, first order ``hyperfine-only'' relaxation is possible. Therefore, resonances in the relaxation rate are to be expected for these crossings. In the very vicinity of such a crossing, electronic states with quite long lifetimes can be expected, if the energy difference between initial and final electronic state is too large for a nuclear absorption (and the hyperfine plus electron-phonon interaction relaxation mechanism is dominant).

Promising for the implementation of a quantum gate, reduced second order relaxation rates have been found for double quantum dots. These rates are much lower than for single dots because of the tunnel splitting. This small energy difference leads to a low phonon density of states entering the second order rate, and thus, to a low transition rate. For asymmetric double dots, this tunneling can be enabled with increasing magnetic fields, possibly leading to a relatively low second order rate for a wide range of magnetic fields.

\bibliography{ddot}

\end{document}